# On the astronomical orientation of the IV dynasty Egyptian pyramids and the dating of the second Giza pyramid.

Giulio Magli
Dipartimento di Matematica, Politecnico di Milano
P.le Leonardo Da Vinci 32, 20133 Milan, Italy
magli@mate.polimi.it

## 1. Introduction: the simultaneous transit method.

It is very well known that the five main pyramids of the fourth dynasty (the main three at Giza and the two Snefru pyramids at Dashur) were oriented with a high degree of precision. It is difficult to fully establish today the original orientation since the casing is almost lost, a part for the so called "bent" Dashur pyramid, the top of the second Giza pyramid, residual blocks in the lowest course of Khufu pyramid, and part of the granite lintel of the third Giza pyramid. However, reasonable data for the orientation can be obtained also from such remains and from accurate measurement of distances between corners.

We shall make use here of the data giving the deviation of the est side from true north, since a complete set of data is avaliable. These are:

Meidum[1] –20' ± 1.0'     [1]
Bent Pyramid -17.3' ± 0.2' ,  [2]
Red Pyramid -8.7' ± 0.2' ,   [2]
Khufu -3.4' ± 0.2' ,      [2]
Khafre -6.0' ± 0.2' .     [2]
Menkaure +12.4' ± 1.0'    []

In recent literature (Spence [3] and Belmonte [4]) also the data for two pyramids of the fifth dynasty (Sahure and Neferirkare) are considered. Such data are taken from the work by Zaba [5] and reported with an error of only 10'. However both buildings are so inferior in work and in a so badly damaged state today, that a measurement of orientation looks difficult and claims of having measured it with a great accuracy look extremely unsound. Actually for the aim of the present paper what might be relevant is the case of the Sahure pyramid which is claimed to be oriented -23' within 10' accuracy. This is, however, absurd since it is well known that the basis of this pyramid was laid down with a displacement of more than 1.5 degrees [6].

Regarding the considered data, although a confidence of only 0.2' (=12 arcseconds) can be questionable in view of the aforementioned difficulties, for the aim of the present paper even a error of one arcminute (certainly exagerate for modern measures) would not invalidate our theses, so that we shall not discuss further the issue of precision of the modern measures. The errors originally made by the builders in identyfing a chosen direction should be assumed of the order of the best naked-eye observation error, namely 1-2 arcminutes. The deviations from north are greater than this value and, in addition, it looks quite strange that their behavior exhibits a minimum at the Great Pyramid. This can be viewed (and will be viewed here) as an experimental datum obtained by a physical measure, which must be explained in physical terms. If we assume, as it is natural to do, that the ability of the builders in checking alignments remained constant in time during the 4 dynasty, it is obvious that there must exist a time-dependent font of systematic error. The unique available phenomenon able to produce this error is, obviously, precession. Thus, as first noticed by Haack [7], all the orientation methods which give precession-independent results all ruled out, including the - otherwise quite sound - method proposed by Edwards consisting in the bisection of the rising and setting positions of a star on a artificial horizon.

---

[1] The datum is somewhat unclear, since Dorner [2] reports a value pratically equal to that of the Bent pyramid (which could imply that the projects started together, see below) while in the literature it is usually reported the datum to which we conform here and due to Petrie.

The problem aimed Kate Spence [3] to propose a new method of orientation which explains the phenomenon including a time-dependent systematic error due to precession. This method – "simultaneous transit" - accounts very well for the observed variation of orientation errors keeping fixed the abilities of the builders, and consists in observing the cord connecting two circumpolar stars, namely Kochab (β UMi) and Mizar (ζ UMa) . When one of the two stars is in the upper culmination the cord is orthogonal to the horizon. Spence proposes this direction to be identified with the north direction sought by the pyramid builders. Due to the precessional motion of earth axis, it is an easy exercise to check that the cord does not identify always the same direction: it actually has a slow movement which brought it from the `left` to the `right` of the pole in the 25 century B.C. Plotting the deviation from north against time, Spence shows that the corresponding straight line fits well with the deviation of the pyramids w.r. to true north if the date of "orientation ceremony" occurred for the Khufu pyramid in 2467 BC ±5y. If one, in turn, accepts the method as the one effectively used, the plot can be used to calibrate the date of construction of all the fourth dynasty pyramids, which turn out to be somewhat 80 years later than usually accepted. No evidence of orientation ceremony exists for the old kingdom pyramids, however the "Stretching of the Cord" foundation ceremony in which the astronomical alignments were fixed is well documented for the orientation of temples and it is actually already present in the old kingdom stela called "Pietra di Palermo" [4].

Further to Spence work, Belmonte [4] proposed that the method actually used consisted in measuring alignments as Spence proposed but using two stars (probably Megrez (δ UMa) and Phecda (γ UMa)) which are not each other opposite to the pole. The pole is thus obtained by elongation of a cord lying below or over it. This looks more natural (at least for modern naked-eye sky-watchers) and reconciles the astronomical chronology with the usually accepted one. However, it should be noted that the astronomical dating of the so called "air shafts" of the Khufu pyramid points rather to support Spence's earlier chronology [8]

## 2. Problems of the "simultaneous transit" method.

The solution proposed by Spence for the distribution of errors in the alignments is not free of problems. These problems can be divided into two groups: methodological and technical.

The methodological problems raise extremely interesting questions, and we deserve to discuss them further in a subsequent work. The point is, anyway, that in Spence's work there is a commixtion between "human-science data" and physical data. In fact, the straight line giving the deviation from true north is produced by a purely physical phenomenon (y-axis) while the disposition of the points on the x-axis is based on archeological-historical (i.e. "human science") data. This is not what occurs in a laboratory: usually an "absolute machine" produces the points on the x-axis and only the data on the y-axis are measured and compared with the theory. Let's make a very trivial example: when we measure activity of a radioactive substance, we choose sample times (a clock gives the points on the x-axis, and we regret errors in time measures as negligible) and we measure activities comparing them with the exponential decay (theoretical) curve. We are, of course, not allowed to change *ad libitum* the disposition of the points on the time axis. In Spence's work, one has to *assume* the validity of the theoretical model and then trace back the "recalibrated" dates on the x-axis. In the author's view, this is an extremely interesting example of the kind of difficulties which have to be expected in archaeoastronomy if one works treating this science as an "exact science" living at the border of "human science" and then tries to remove the border. In the particular case under examination, the best way to solve the problem would be finding an independent way of relative dating between the monuments. It is obvious, that carbon-dating can be of no help here, but there are some chances of high precision dating of the fourth dynasty pyramids via dendrochronology which seem to have been overlooked in the past, since - as far as the present author is aware - neither the cedar beams of the burial chamber of the Bent pyramid nor the wood of Khufu' boats have been dated in this way. In addition it seems that an original wood beam is also present in the lower northern shaft of Khufu [9].

We now turn to technical problems. There are two: the "Abu Roash" problem and the "inverse point" problem.

Djedefre, son of Khufu, ruled between Khufu and Khafre. He started building his pyramid at Abu Roash. This pyramid was left incomplete, and only the substructure with the descending passage and the first courses where probably constructed, although it is difficult to know how much of the building was destroyed in the past. When papers [3,4] where published, the data of the orientation of the Abu Roash pyramid where not available. Such data are available today[2] [10] and *do not*, by all means, fit in the "simultaneous transit" framework. In fact the alignment error turns out to be -48.7' (the error on this measure is very small, a fraction of arcminute).
This is clearly a problem for the whole framework, and it may well be that it simply shows that the theory is incorrect (that is, the simultaneous transit method was not used). We do not have the answer. However, we would like to make the following comment. An error of about 5/6 degree is *orribly huge* for the IV dinasty standards. It is near one order of magnitude greater than those made for Khufu a few years before and those made for Khafre a few years later. Thus, one is really tempted to consider Abu Roash as an exception, a case in which, for some unknown reason, perhaps ritual, a different and much more imprecise method (such as a solar shadow method) was used, and we shall proceed further using this assumption. However, we stress again that this interpretation has some chance exactly because the datum does not fit in the framework in a clamorous way.

We now turn to the "inverse point" problem. It consists in the fact that the orientation of the Khafre pyramid fits in the calibration line if and only if the corresponding point is "lifted up" vertically in the positive region. This of course is a serious objection to the method. To solve the problem, Spence [3] speculates that the orientation of the second pyramid was carried out in the opposite season with respect to the others. It is obvious, that in this case the cord between the two stars has the opposite position with respect to the pole it would have the same year in the opposite season, and so the "flipped" data fit again. Spence thus proposes that Khafre pyramid was the unique to be oriented in the summer season, all the other pyramids being instead oriented in winter. Also in the Belmonte [4] proposal the problem has to be solved by assuming a special procedure for the orientation of the Khafre pyramid. In this case, one has to assume that this pyramid was the unique one to be oriented with the two proposed stars in lower culmination, all other pyramids being oriented with the stars in the upper culmination.

---

[2] The author is grateful to John Wall for signalling this reference.

## 3. An inverse chronology at Giza?

It is clear that the above mentioned exceptionality of the second pyramid is ` embarrassing` . In the case of Spence' s theory, it is rather strange that a procedure which should by any evidence be intended as being of religious and deep nature, such as the orientation of a giant king' s tomb, could occur scattered in time rather then in a fixed, precise time dictated by astronomical counting, such as e.g. those rituals connected with the Sirius cycle. It is, therefore, difficult to agree with Spence' s ingenious explanation for the "lifting up" of the Khafre point on the calibration line due to inverse season orientation. In the case of Belmonte' s proposal, it is clear (and it is, in fact, thoroughly discussed by the same author) that it is rather simple and natural the use of the lower culmination rather than the upper one. For instance, in the case of the lower culmination an instrument such as the very well documented Merkhet could be used (due to the relatively small angle formed by the alignment and the horizon) while for a measure based on the upper culmination a new sort of instrument must be supposed (see [4] for details).

There is a simple, natural, nearly obvious it physical interpretation of the data which leads to a simple solution for the problem of the "wrong sign" of Khafre' s orientation. The error in the orientation of the second pyramid would fit perfectly well if the point representing Khafre is *translated orizzontally* until it reaches the calibration line. As a result however, the point on the time axis which gives the date of construction of the second pyramid now occurs before that of the first pyramid.[3]

Let us review which kind of evidence is available regarding the fact that the second Giza pyramid was constructed after the first. Is such an assumption well grounded on sound archeological data? Leaving aside the 2000 y later attribution by Herodotus of the Giza pyramids, the first pyramid is soundly attributed to Khufu due to the so called "quarry marks" discovered in the upper four of the five so called "relieving chambers" of the King chamber, while no inscription, of any kind and nature, mentioning Khafre or any other fourth dynasty king and dating to the same period has ever been found in the whole second pyramid complex namely the pyramid, the causeway, the valley temple, the Sphinx temple, the Sphinx itself. The unique available proof consists in the discovering of the famous Khafre diorite statue in the valley temple and fragments of other statues of the same ruler (the very debated presence of Khafre name as the owner of the sphinx in the "dream stela" is anyway on a 1000 years later inscription). However the presence of the statue cannot be considered as a definitive proof of ownership, taking also into account that "robbing" of monuments was rather common in ancient Egypt.

---

[3] It is worth noticing that, if the datum for the Sahure pyramid (-21') given by Zaba and reported by Spence is taken seriuously then our proposal has no chance to work, since the same procedure applied to it would lead to a foolish, earlier datation for this pyramid. However,as already noticed, this datum cannot be considered as correct.

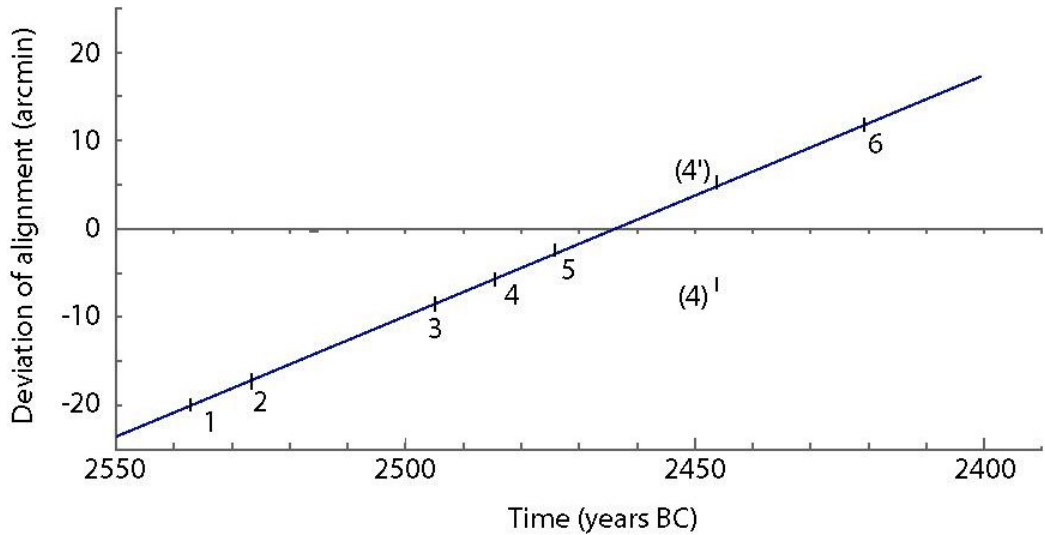

Figure 1
The solid line is a plot of the deviation from true north (in arcminutes) of the cord between Kochab and Mizar vs. time. If the method used for the alignment of the pyramids is that proposed by Spence, than the numbered points allow a recalibration of the dating: 1.Meidum 2.Bent 3.Red 4.Khafre 5.Khufu 6.Menkaure. The second pyramid here occurs before Khufu for the reasons explained in the text. The point (4) represents the same pyramid in the standard relative cronology while the point (4') represents again the same pyramid but `flipped` in the positive region in accordance to Spence's `inverse season` proposal.

On the other side, there are hints pointing to a pre-building of the second pyramid with respect to Khufu pyramid. First of all, everybody visiting Giza knows that, although being slightly smaller than Khufu, the Khafre pyramid looks greater because it was constructed on a more elevated position. Why did not Khufu choose such a position, being the first? Why Khafre choose to build a smaller pyramid, but not a much smaller one, justifiable by economical reasons? Michelangelo in building the S. Peter cupola in Rome stated "Piu' grande ma non piu' bella" ("greater, but not more beautiful") referring to the Brunelleschi cupola of St. Maria del Fiore in Florence.

The idea that ``Khafre could precede Khufu'' is not in itself new and has been proposed by various authors before, on the basis however of different motivations (see e.g. [11]). Actually, there are geological evidences showing that the construction of the Khafre causeway took place before that of the Khufu pyramid, since the causeway strangely seems to delimitate one of the borders of the Khufu quarry [12]. If we accept for a moment that the second pyramid was the first to be constructed the astronomical data suggest a period very close to the year 2500 b.c. (close to 2580 B.C. if Belmonte's method is adopted). If this pyramid was really built before Khufu's one, then it is necessary to understand who was the builder. One could however also speculate that it was Khufu himself to build two pyramids as probably Snefru did at Dashur. It is, in fact, not very clear the reason for which Snefru was the builder of two pyramids, and this reason need not necessarily be of technical nature. There are evidences of structural problems in the Bent pyramid, consisting essentially in a dislocation of the exterior mantle of core blocks in the whole monument, but it is far from clear that such problems arised during construction, leading to the decision of varying the angle first and then building a new pyramid [13].

It is obvious, that our proposal can have a chance to work only if Khafre's tomb an be accomodated in the picture. In this respect there are two possibilities. First of all, there is no clear evidence for the attribution of the IV dynasty unfinished pyramid of Zawyet el Arian which thus could be assigned to Khafre. More simply, one could suppose that Khufu started two projects and then Khafre picked up one of them, since there are evidences that the second pyramid was fully attributed to this ruler already in the VI dinasty.[4]

All in all, since there is no striking evidence for the standard chronology between the two pyramids, the *possibility* of an inverse chronology suggested by purely astronomical data cannot be rejected on available purely archeological data. There are some possible research directions. One is the abovementioned possibility of dendrochronological dating of the wood remains. From the archaeo-astronomical point of view it would be of extreme interest comparing the data on the orientation of the Sphinx and of the Khafre Valley Temple with those of the pyramids, as well as obtaining data from the Zawyet el Arian site (the author has been unable to find such data in the literature so far). Definitive evidence could also come from rigorous geological analysis of the quarries at Giza, which waits since many years to be carried out in a systematic way.

On the other side, the "out of scale" datum coming from Abu Roash remains as an overall problem for all the theories based on the simultaneous transit method.

## Acknowledgements

A warm thank to J.A. Belmonte for a careful reading of the first version of this work and his constructive comments. Comments by R.G. Bauval are also gratefully acknowledged.

---

[4] I'm grateful to J. A. Belmonte for pointing out this fact.